# Interaction of an acoustical quasi-Gaussian beam with a rigid sphere: linear axial scattering, instantaneous and time-averaged radiation force

F.G. Mitri[†], *Member IEEE*

[†]Los Alamos National Laboratory (LANL), MPA-11, Sensors & Electrochemical Devices, Acoustics & Sensors Technology Team, MS D429, Los Alamos, NM 87545, USA

Corresponding author:

F.G. Mitri, emails: mitri@ieee.org, mitri@lanl.gov

*Abstract* – This work focuses on the interaction of an acoustical quasi-Gaussian beam centered on a rigid immovable sphere, during which at least three physical phenomena arise, namely, the (axial) acoustic scattering, the instantaneous force, and the time-average radiation force which are investigated here. The quasi-Gaussian beam is an exact solution of the source free Helmholtz wave equation and is characterized by an arbitrary waist $w_0$ and a diffraction convergence length known as the Rayleigh range $z_R$. Specialized formulations for the scattering and the instantaneous force function as well as the (time-averaged) radiation force function are provided. Numerical computations illustrate the variations of the backscattering form-function, the *instantaneous* force function and the (time-averaged) radiation force function versus the dimensionless frequency *ka* (where *k* is the wave number and *a* is the radius of the sphere), and the results show significant differences from the plane wave limit when the dimensionless beam waist parameter $kw_0 < 25$. The radiation force function may be used to calibrate high-frequency transducers operating with this type of beam. Furthermore, the theoretical analysis can be readily extended to the case of other types of spheres (i.e. elastic, viscoelastic, shells, coated spheres and shells) providing that their appropriate scattering coefficients are used.

# I. INTRODUCTION

Gaussian beams have been shown to be of particular interests in applications involving the acoustic scattering of finite beams by a sphere [1], the time-averaged acoustic radiation force in acoustical tweezers and particle entrapment [2], acoustic microscopy [3, 4], medical diagnosis [5, 6], and nondestructive evaluation [7] and imaging [8, 9] of materials to name a few.

Gaussian beams originate in the wave diffraction theory as a solution of the parabolic wave equation, which describes the pressure field in the paraxial approximation only. When the beam waist radius is larger than the acoustic wavelength, the scattering properties of such beams can be analyzed [1] based on the solution that satisfies the source free Helmholtz equation.

Despite the fact that Gaussian beams do not generally satisfy the Helmholtz equation, there exist *beams'* expressions that are exact solutions of the source free Helmholtz wave equation, such as Bessel beams [10, 11] and their superposition [12, 13], so that their linear scattering [12, 14-20] and acoustic forces [21-29] properties can be analyzed using the standard potential field technique combined with a standard wave-decomposition method.

Recently, a new family of beams that satisfies the source free Helmholtz equation has been introduced; these are known as quasi-Gaussian beams [30]. A quasi-Gaussian beam is characterized by an arbitrary waist $w_0$ and a diffraction convergence length known as the Rayleigh range $z_R$. Moreover, the beam has the form of a superposition of sources and sinks with complex coordinates [30].

The properties of such a beam have been analyzed from the standpoint of the classical wave propagation theory [30], however, the solutions for the scattering and acoustic forces have not been provided yet. It is the aim of the present work to provide such solutions which are important in applications involving particle manipulation, calibration of ultrasound transducers, and imaging using this new type of acoustical beam. Here, the physical phenomena related to the acoustic scattering and the instantaneous force as well as the time-averaged radiation force resulting from the interaction of a quasi-Gaussian beam are investigated. The problem of the scattering of a quasi-Gaussian beam is exactly soluble; this is done for a rigid immovable sphere mimicking a liquid drop in air (although the analysis may be readily extended to the case of an elastic or viscoelastic sphere, or spherical shell, bare or coated with an absorbing layer) and the scattering solution is used to derive the instantaneous and time-averaged forces. The example for the rigid immovable sphere has some relevance in fluid dynamics applications because it simulates a liquid drop (considered as a sphere) in air. Since the drop is usually about 813 times denser than air and the acoustic impedances of liquid and air are mismatched, liquid drops in air are approximated as perfectly rigid objects, and the scattering and acoustical forces are not sensitive to the density or sound speed of the drop.

# II. ACOUSTICAL SCATTERING OF A QUASI-GAUSSIAN BEAM BY A SPHERE

Consider an incident acoustical wave field in the form of a quasi-Gaussian beam *centered* on a rigid immovable sphere of radius $a$ and described by its complex pressure $P_i$ that is a solution of the Helmholtz wave equation,

$$\left(\nabla^2 + k^2\right) P_i = 0, \tag{1}$$

where $k$ is the wave number.

In a system of spherical coordinates ($r$, $\theta$, $\phi$) with its origin chosen at the center of the sphere, the incident pressure field solution of Eq.(1), is expressed as [30],

$$P_i(r,\theta,\phi) = P_0 \sum_{n=0}^{\infty} i^n (2n+1) g_n(kz_R) j_n(kr) P_n(\cos\theta), \tag{2}$$

where $P_0$ is the amplitude, $j_n(.)$ is the spherical Bessel function of the first kind, $P_n(.)$ are Legendre functions, $z_R$ is the Rayleigh range in the axial direction of wave propagation that is given by,

$$z_R = \frac{kw_0^2}{2}, \tag{3}$$

where $w_0$ is the beam's waist (denoted by "$a$" in [30]), and the (real) coefficient $g_n$ in Eq.(2) is given by,

$$g_n(kz_R) = \frac{\left[1-(-1)^n e^{-2kz_R}\right]}{\left(1-e^{-2kz_R}\right)^2} e^{-kz_R} \sqrt{2\pi kz_R} I_{n+1/2}(kz_R), \tag{4}$$

where $I_{n+1/2}(\cdot)$ is the modified cylindrical Bessel function of the first kind [31].

It has been noticed that for wide beams $(kw_0 \gg 1)$ and $kz_R \to \infty$, the coefficients $g_n(kz_R) \to 1$ [30], and Eq.(2) reduces to the case of plane progressive waves [32].

The presence of the sphere in the waves' path causes the incident pressure field to scatter. The pressure scattered by the sphere produces a spherical wave-field which is represented by

$$P_s(r,\theta,\phi) = P_0 \sum_{n=0}^{\infty} i^n (2n+1) g_n(kz_R) S_n(ka) h_n^{(1)}(kr) P_n(\cos\theta), \qquad (5)$$

where $h_n^{(1)}(.)$ is the spherical Hankel function of the first kind, and $S_n(ka)$ are the scattering partial-wave coefficients of the sphere. For a rigid immovable sphere, these coefficients are determined using the solution that satisfies the boundary condition of the vanishing of the particle velocity at the boundary $r = a$ [33] such that,

$$S_n(ka) = -\frac{j_n'(ka)}{h_n^{(1)'}(ka)}. \qquad (6)$$

It is common to investigate the acoustic scattering in the far-field region. Therefore, in the far-field region $(kr \to \infty)$, the steady-state (time-independent) scattered pressure from a spherical target given by Eq.(3) can be expressed as [34, 35]

$$P_s(r,\theta,\phi) \underset{kr \to \infty}{=} P_0 \frac{a}{2r} f_\infty(ka, kz_R, \theta) e^{ikr}, \qquad (7)$$

where the spherical Hankel function of the first kind reduces to the following asymptotic approximation; $h_n^{(1)}(kr) \xrightarrow[kr \to \infty]{} \frac{1}{i^{(n+1)} kr} e^{ikr}$.

The form function for a spherical shell $f_\infty$, is therefore defined by the partial wave series as,

$$f_\infty(ka, kz_R, \theta) = \frac{2}{ika} \sum_{n=0}^{\infty} (2n+1) g_n(kz_R) S_n(ka) P_n(\cos\theta). \tag{8}$$

## III. INSTANTANEOUS FORCE OF A QUASI-GAUSSIAN BEAM ON A SPHERE

Another phenomenon that arises from the interaction of an acoustical quasi-Gaussian beam with a sphere is the *transient* (instantaneous) force.

The general calculation of the *instantaneous* force is obtained by integrating the total (incident + scattered) linear pressure (p. 353, in [36]) generated by the quasi-Gaussian sound field over the time-varying (vibrating) surface of the sphere $S(t)$. The general formula is given by (p. 2, §2 in [37]),

$$\mathbf{F}_{inst} = \text{Re}\left[ -\iint_{S(t)} \left( P_i(r,\theta,\phi) + P_s(r,\theta,\phi) \right) e^{-i\omega t} \mathbf{n}\, dS \right], \tag{9}$$

where $\mathbf{n}$ is the outward normal to the surface of the sphere, and $dS = r^2 \sin\theta\, d\theta\, d\phi$ is the elementary surface element.

For a sphere centered on the beam's axis, symmetry requires that the force has only an axial component. In spherical coordinates, the instantaneous force is expressed as,

$$\mathbf{F}_{inst}\big|_{r=a} = \text{Re}\left[ \hat{F}_{z,inst}\big|_{r=a} e^{-i\omega t} \mathbf{e_z} \right] = \text{Re}\left[ -a^2 \int_0^{2\pi} \left( \int_0^{\pi} \left( P_i(r,\theta,\phi) + P_s(r,\theta,\phi) \right) \cos\theta \sin\theta\, d\theta \right) d\phi\, e^{-i\omega t} \mathbf{e_z} \right], \tag{10}$$

where $\mathbf{e}_z$ is the unitary vector along the axial direction. It is important to note here that integration of the pressure is performed over the surface of the sphere *at rest* because it is rigid and immovable.

Substituting Eqs. (2) and (5) into Eq.(10) after denoting the finite value of the integral, $I_1 = \int_0^\pi P_n(\cos\theta)\cos\theta\sin\theta\,d\theta = \frac{2}{3}\delta_{n1}$, where the symbol $\delta_{ij}$ denotes the Kronecker delta, the complex instantaneous force amplitude $\hat{F}_{z,p}$ is expressed as,

$$\hat{F}_{z,inst}\Big|_{r=a} = -4i\pi a^2 P_0 \left[ j_1(ka) + A_1 h_1^{(1)}(ka) \right] g_1(kz_R). \tag{11}$$

It is noticeable that only the dipole partial-wave ($n = 1$) contributes to the instantaneous force because of the property of the integral $I_1$.

## IV. TIME-AVERAGED RADIATION FORCE OF A QUASI-GAUSSIAN BEAM ON A SPHERE

When the beam is composed of continuous waves, the sphere experiences a steady radiation force, which is calculated by integrating the time-averaged excess of pressure (taken up to second-order quantities) over the sphere's surface at rest as [38],

$$\langle \mathbf{F}_{rad} \rangle = \left\langle \iint_{S_0} \mathscr{L} \mathbf{n}\, dS \right\rangle - \left\langle \iint_{S_0} \rho \mathbf{v}^{(1)} \left( \mathbf{v}^{(1)} \cdot \mathbf{n} \right) dS \right\rangle, \tag{12}$$

where,

$$\begin{aligned}\mathscr{L} &= \frac{\rho}{2}\left|\mathbf{v}^{(1)}\right|^2 - \frac{1}{2\rho c^2}\left|p^{(1)}\right|^2, \\ &= \mathscr{K} - \mathscr{U},\end{aligned} \tag{13}$$

is the Lagrangean energy density, the superscript $^{(1)}$ denotes first-order quantities, $\mathbf{v}^{(1)} = \nabla \varphi$, $p^{(1)} = \text{Re}\left[\left(P_i(r,\theta,\phi) + P_s(r,\theta,\phi)\right)e^{-i\omega t}\right] = -\rho \frac{\partial \varphi}{\partial t}$, and $\varphi = \text{Re}[\Phi]$, where $\Phi$ is the total (incident + scattered) linear velocity potential that is related to the total pressure. It is important to note the fundamental distinction [39-41] between the "instantaneous" force due to a transient acoustical wave (treated in Section III) and the terminology of the steady-state (or static) radiation force of a beam composed of continuous waves and treated in this section.

For a rigid immovable sphere, at the interface fluid-rigid surface, there is absence of oscillatory movement. Hence, the normal component of the fluid particle velocity $v_n = \mathbf{v}^{(1)} \cdot \mathbf{n}$ vanishes, and Eq.(12) reduces to,

$$\langle \mathbf{F}_{rad} \rangle = \left\langle \iint_{S_0} \mathscr{L} \mathbf{n} \, dS \right\rangle. \tag{14}$$

In the direction of wave propagation (i.e. the axial $z$-direction), the radiation force on the sphere can be therefore expressed as

$$\begin{aligned}\langle F_{z,rad} \rangle &= \langle \mathbf{F}_{rad} \rangle \cdot \mathbf{e_z}, \\ &= \langle F_r \rangle + \langle F_\theta \rangle + \langle F_{r,\theta} \rangle + \langle F_t \rangle,\end{aligned} \tag{15}$$

where,

$$\langle F_r \rangle = (-a^2\rho/2)\left\langle \int_0^{2\pi}\left\{\int_0^\pi [\partial_r\varphi]_{r=a}^2 \sin\theta\cos\theta\,d\theta\right\}d\phi\right\rangle,$$

$$\langle F_\theta \rangle = (\rho/2)\left\langle \int_0^{2\pi}\left\{\int_0^\pi [\partial_\theta\varphi]_{r=a}^2 \sin\theta\cos\theta\,d\theta\right\}d\phi\right\rangle,$$

$$\langle F_{r,\theta} \rangle = (a\rho)\left\langle \int_0^{2\pi}\left\{\int_0^\pi [\partial_r\varphi]_{r=a}[\partial_\theta\varphi]_{r=a} \sin^2\theta\,d\theta\right\}d\phi\right\rangle, \quad (16)$$

$$\langle F_t \rangle = (-a^2\rho/2c^2)\left\langle \int_0^{2\pi}\left\{\int_0^\pi [\partial_t\varphi]_{r=a}^2 \sin\theta\cos\theta\,d\theta\right\}d\phi\right\rangle,$$

and $\varphi$ is rewritten as,

$$\varphi = \Phi_0 \sum_{n=0}^\infty (2n+1) R_n P_n(\cos\theta), \quad (17)$$

where, $\Phi_0$ is the (real) amplitude,

$$R_n = \mathrm{Re}\left[i^n\left(U_n(kr)+iV_n(kr)\right)g_n(kz_R)e^{-i\omega t}\right], \quad (18)$$

and

$$\begin{aligned} U_n &= (1+\alpha_n)j_n(kr) - \beta_n y_n(kr), \\ V_n &= \beta_n j_n(kr) + \alpha_n y_n(kr), \end{aligned} \quad (19)$$

where $y_n(.)$ are the spherical Neumann functions (or the spherical Bessel functions of the second kind), $\alpha_n = \mathrm{Re}[S_n]$, and $\beta_n = \mathrm{Im}[S_n]$.

Substituting Eq.(17) into Eq.(15) using Eqs.(16), (18) and (19), manipulating the results, and denoting by $E = \rho k^2 \Phi_0^2/2$ the characteristic energy density, the final expression for the axial time-averaged radiation force of a quasi-Gaussian beam given by Eq.(15) is simplified and is expressed by,

$$\langle F_{z,rad} \rangle = Y_{qG} S_c E, \quad (20)$$

where $S_c = \pi a^2$ is the cross-sectional area. The dimensionless factor $Y_{qG}$ is given by,

$$Y_{qG} = -\frac{4}{(ka)^2} \sum_{n=0}^{\infty} \left\{ g_n(kz_R) g_{n+1}(kz_R)(n+1)\left[\alpha_n + \alpha_{n+1} + 2(\alpha_n \alpha_{n+1} + \beta_n \beta_{n+1})\right] \right\}. \tag{21}$$

This reduces to the expression of $Y_p$ given previously [42, 43] in the limit of plane waves (i.e. $g_n(kz_R) \to 1$).

## V. NUMERICAL RESULTS AND DISCUSSION

To illustrate the interaction of an acoustical quasi-Gaussian beam with a rigid immovable sphere, numerical simulations are performed for the magnitude of the backscattering form-function $|f_\infty(ka, kz_R, \pi)|$, the magnitude of the instantaneous force function $|Y_{qG,inst}| = |\hat{F}_{z,inst}|_{r=a} / \pi a^2 P$, and the radiation force function $Y_{qG}$ which is the radiation force per unit characteristic energy density and unit cross-sectional area.

The simulations are evaluated in the dimensionless frequency range $0 < ka \leq 10$ for chosen values of the dimensionless beam waist parameter $kw_0$ (= 1, 3, 5, 10, 25, $\to\infty$), respectively, where the case $kw_0 \to \infty$ corresponds to plane waves.

Figure 1 shows the plots for the backscattering form-function for a rigid immovable sphere for various dimensionless beam waist values. As the parameter $kw_0$ increases from unity, the curves for the backscattering form-function show fluctuations as the plane wave limit is reached for $kw_0 > 25$. Such fluctuations appear to be highly dependent on the value of $kw_0$.

In Fig. 2, the plots for the magnitude of the instantaneous force function $|Y_{qG,inst}|$ are displayed versus $ka$ and $kw_0$. For $ka < 1.4$, all the $|Y_{qG,inst}|$ plots linearly increases versus *ka* to reach their maxima at *ka* = 1.421, but beyond this limit (i.e., when the wavelength becomes smaller that the circumference of the sphere), the linear relation breaks down and $|Y_{qG,inst}|$ diminishes with increasing *ka* values. In addition, as $kw_0$ increases beyond 25 and tends to infinity, $|Y_{qG,inst}|$ increases and approaches the plane wave limit.

Figure 3 displays the plots for the radiation force function $Y_{qG}$ versus *ka* and $kw_0$. As observed, the radiation force function $Y_{qG}$ increases with increasing $kw_0$ to reach the plane wave limit [43] when $kw_0 \to \infty$. Significant differences are observed between the quasi-Gaussian beam and the plane wave results for $kw_0 < 25$. A similar behavior has been observed for a sphere in the field of a focused transducer [44].

It is important to note that predictions of the $Y_{qG}$ using this theory allow the calibration of the yet to developed quasi-Gaussian sonic probes from radiation force measurements, and expands the utility of the sphere radiometer method [44, 45]. Additional applications may be sought in particle manipulation research using this new type of beams since the theoretical computations give *a priori* information on the force used to manipulate, or levitate a spherical particle in a host fluid.

## VI. CONCLUSION

In this work, the scattering and radiating force properties of a quasi-Gaussian beam incident upon a rigid, immovable sphere are investigated. The quasi-Gaussian beam represents an exact solution of the source-free Helmholtz equation and is characterized by an arbitrary waist $w_0$ and a diffraction convergence length known as the Rayleigh range $z_R$. Analytical (closed-form) expressions under the form of partial-wave series are presented and representative plots describing the backscattering form function, the instantaneous force as well as the time-averaged radiation force are provided, with particular emphasis on the dimensionless beam waist parameter $kw_0$ and the dimensionless frequency *ka*. It is shown that significant differences are observed between the quasi-Gaussian beam and the plane wave results for $kw_0 < 25$, and the plane wave results are recovered when $kw_0 \to \infty$.

**Acknowledgments:** The financial support provided through a Director's fellowship (LDRD-X9N9, Project # 20100595PRD1) from Los Alamos National Laboratory is gratefully acknowledged. Disclosure: this unclassified publication, with the following reference no. LA-UR 12-22129, has been approved for unlimited public release under DUSA ENSCI.

# Figures

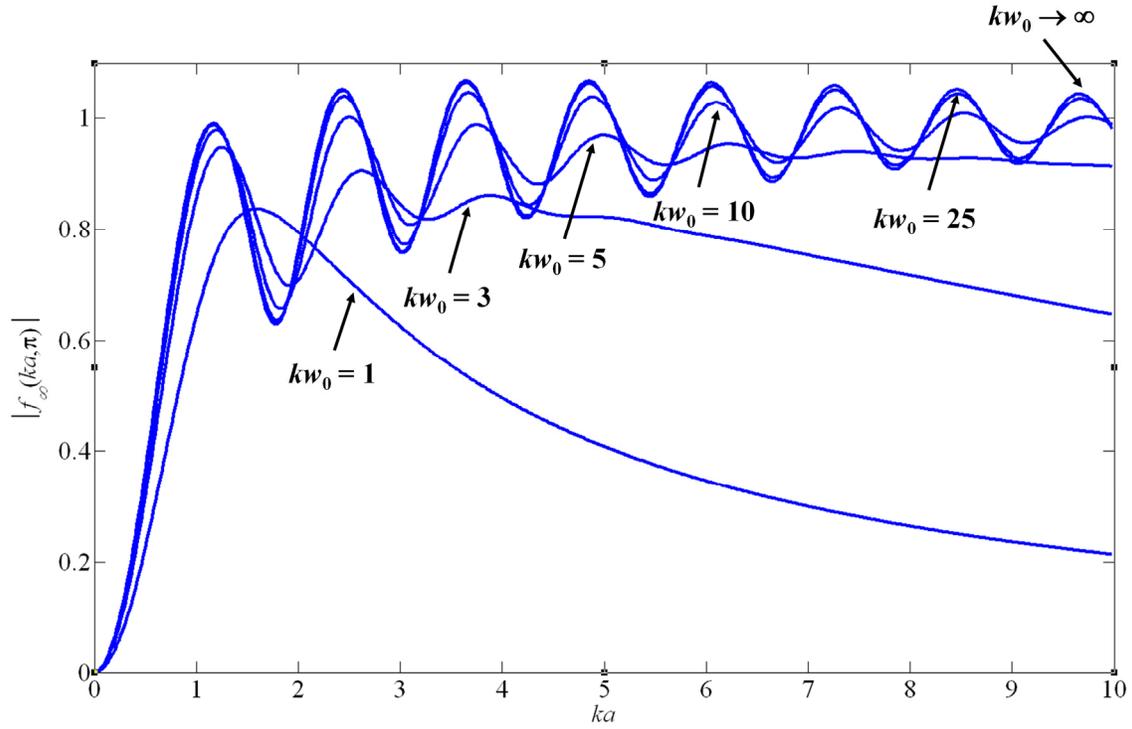

**Fig. 1.** Plots for the magnitude of the backscattering form-function for a quasi-Gaussian beam incident upon a rigid immovable sphere. The case where $kw_0 \to \infty$ corresponds to plane waves.

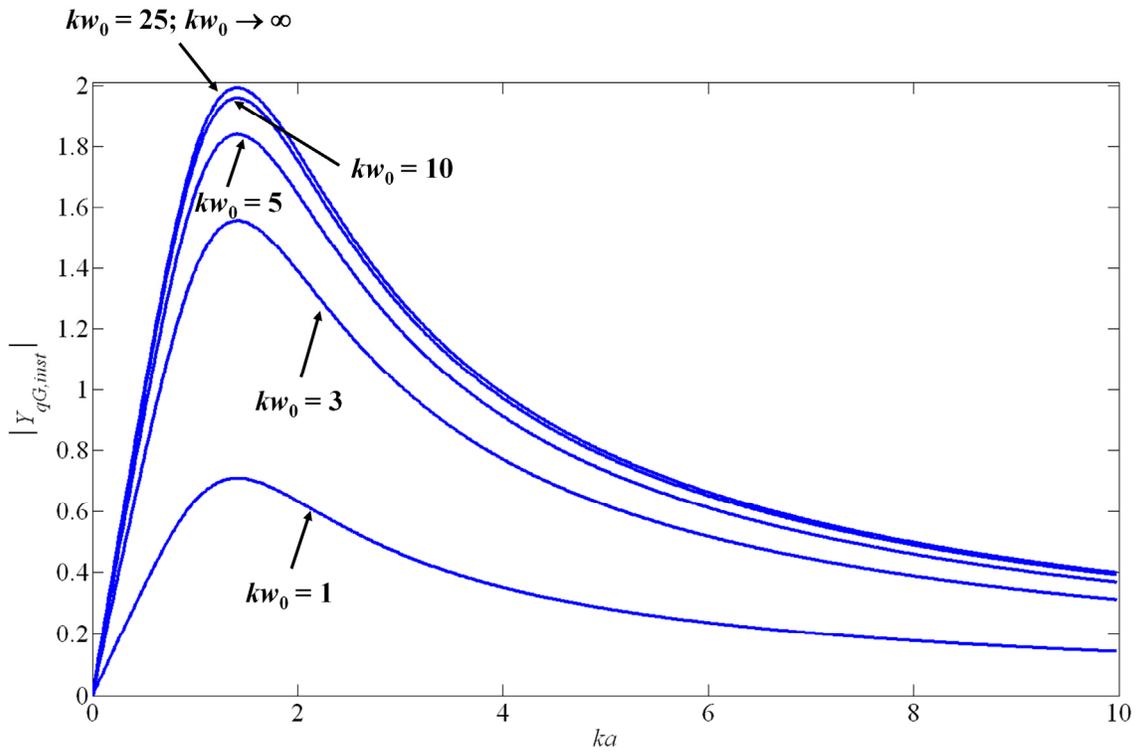

**Fig. 2.** The same as in Fig. 1 but the plots correspond to the magnitude of the instantaneous force function that is displayed for various values of $kw_0$.

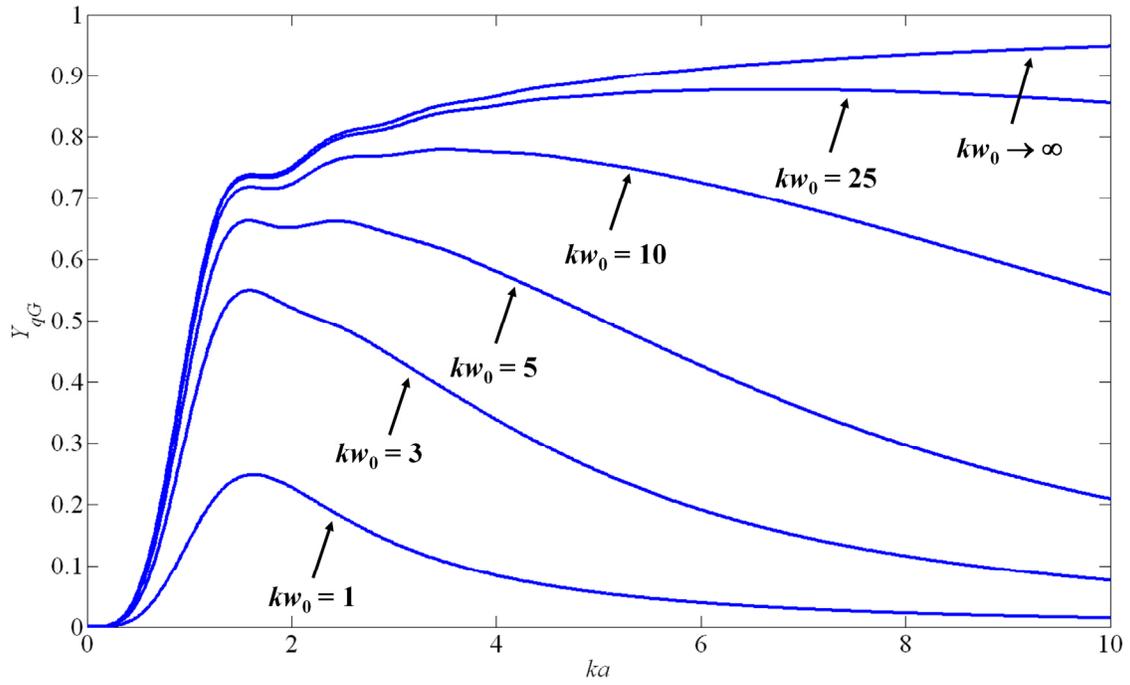

**Fig. 3.** The same as in Fig. 2, but the plots correspond to the radiation force function for various values of $kw_0$.